# A comparative nanotribological investigation on amorphous and polycrystalline forms of $MoS_2$

Hesam Khaksar[1(*)], Prashant Mittal[2(*)], Nabil Daghbouj[2], Grzegorz Cios[3], Tomas Polcar[2], and Enrico Gnecco[1]

[1]*Marian Smoluchowski Institute of Physics, Jagiellonian University, 30-348 Krakow, Poland*

[2]*Department of Control Engineering, Faculty of Electrical Engineering, Czech Technical University in Prague, Prague, Czech Republic*

[3] *AGH University of Science and Technology, Academic Centre for Materials and Nanotechnology, Al. A. Mickiewicza 30, 30-059 Krakow, Poland*

[(*)]*Both authors equally contributed to this work*

## Highlights

- The wear response of amorphous and polycrystalline forms of MoS2 is compared on the nanoscale
- Archard's wear equation is applicable and the wear resistance is four times higher in the amorphous form.
- The coefficient of friction is significantly reduced by the surface smoothing induced by the wear process

## Abstract

The wear behavior of two amorphous and polycrystalline forms of $MoS_2$ prepared by magnetron sputtering has been characterized in a combined nanoindentation and atomic force microscopy study. From the analysis of the depth and width of wear tracks estimated after scratching the surfaces with a Berkovich indenter and a loading force up to 2 mN, we conclude that both forms follow the Archard wear equation, and the wear resistance is about four times higher on the amorphous $MoS_2$. Moreover, a comparison of lateral force maps on pristine and worn areas shows a considerable reduction of friction on both forms, which is possibly due to the significant smoothing of the surfaces caused by scratching. With normal forces in the μN range, the analysis is made difficult by the fact that the linear dimensions of the wear tracks are comparable to those of the granular structures forming the surfaces. Even if the Archard equation could not be tested in this case, the wear resistance is considerably larger on amorphous $MoS_2$ also on the nanoscale. In this way, our results disclose information on the nanotribology of $MoS_2$ thin films in forms different from the layered structures commonly discussed in the literature. The amorphous form outperforms the polycrystalline one.

## Graphical abstract

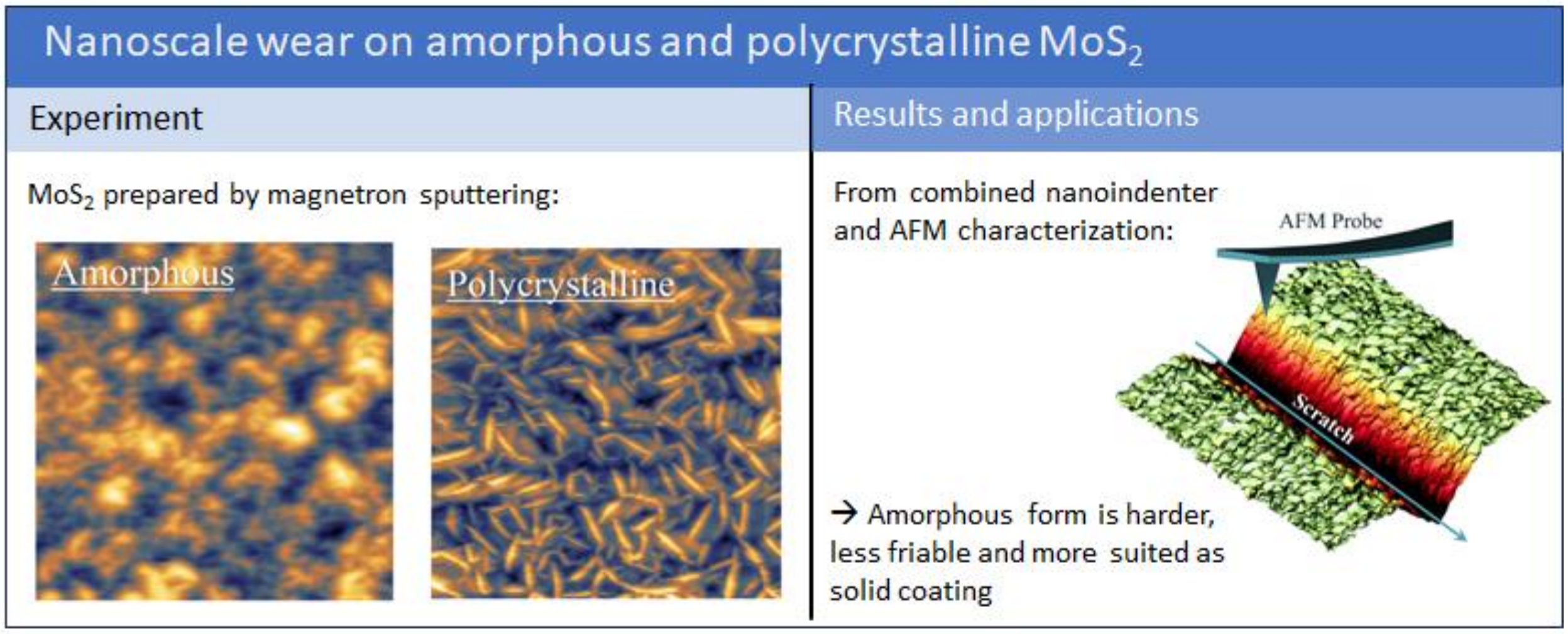

## 1. Introduction

Molybdenum disulfide ($MoS_2$) is widely used as a dry lubricant in extreme environments due to its enhanced performance in the lack of oxygen and its stability up to temperatures of several hundred °C [1]. For these reasons, $MoS_2$ has well-established applications in the automotive [2] and aerospace industry [3], metal forming and cutting tools [4], electrical contacts, and nano- and micro-electromechanical systems [5]. In this context, it is well-known that crystalline forms of $MoS_2$ have better frictional properties as compared to amorphous ones. They are also less sensitive to humidity and atmospheric conditions [6]. However, crystalline $MoS_2$ tend to exfoliate during sliding if the loading forces are too high, resulting in high wear, COF, and low stability [1]. To improve design and durability of components based on $MoS_2$ it just becomes of utmost importance to explore the tribological behavior of $MoS_2$ in alternative forms, and to characterize their early-stage wear behavior down to the nanoscale.

On crystalline $MoS_2$, molecular dynamics (MD) simulations have been helpful in estimating specific damage thresholds for mono and bilayer $MoS_2$ films [7, 8]. However, predicting the performance of $MoS_2$ using MD is challenging for the amorphous and polycrystalline coatings produced by sputtering that are commonly used as lubricants in extreme environments. In this case experimental characterizations such as those made possible by atomic force microscopy (AFM) are the most important tool to gain quantitative insights on the wear resistance of those materials on the nanoscale. As an example, on monolayer $MoS_2$ scratched along its crystallographic armchair direction, AFM has allowed to observe exfoliations at ±30° (i.e. along the zigzag direction) all along the wear tracks, which ended in the folding of flat triangular flakes [9]. Wrinkling and multiple folding was also found as a possible consequence of rippling in the underlying ($SiO_2$) substrate, which is also scratched by the AFM tip. On multilayer $MoS_2$, AFM-based nanoscratching resulted in thicker chips of about 100 nm in length and a height ranging up to a few tens of nm. The present work aims to extend such a characterization of nanowear processes to amorphous and polycrystalline $MoS_2$. To ensure a comprehensive examination and explore a broader range of loading conditions, AFM is combined with nanoindentation based on a Berkovich tip.

## 2. Materials and methods

### 2.1. Sample Preparation

$MoS_2$ films were deposited on polished Si (100) wafer in DC magnetron sputtering configuration, using AJA Orion 5 plasma vapor deposition machine. Specifically, a $MoS_2$ target (diameter of 2”, height of 0.25”, 99.5 % purity) was DC magnetron sputtered in an argon gas atmosphere at 400 °C and at room temperature to prepare polycrystalline and amorphous $MoS_2$, respectively. The Si substrates were sputter-cleaned in the Ar atmosphere at 5 mTorr for 20 minutes. For all depositions, the $MoS_2$ target and plasma source powers were held constant at 80 W. The target power was ramped up at the beginning and ramped down at the end of the process. The substrate bias was 6 W (corresponding to –40 V of substrate potential), and the total working pressure was kept constant at 3 mTorr (0.4 Pa approx.). The total deposition time for the $MoS_2$ layer varied from 120 to 150 minutes (depending on the deposition rate) to achieve a thickness of approximately 1.1 μm for amorphous $MoS_2$ (Fig. 1a) and 2.4 μm for polycrystalline $MoS_2$ (Fig. 1b) at room and high temperatures. Figure 1c shows the Raman spectra of the amorphous and polycrystalline form of $MoS_2$ so obtained. It can be observed that amorphous $MoS_2$ exhibits broader peaks as compared to polycrystalline $MoS_2$. This is due to lack of long-range order which indicates the disordered nature of the amorphous material. Characteristic Raman peaks in polycrystalline $MoS_2$ include the $E^1_{2g}$ mode at approximately 380-400 $cm^{-1}$ and the $A_{1g}$ mode at approximately 400-420 $cm^{-1}$. These peaks reveal details about the symmetry and structure of the crystal and are produced by the vibrational modes of the crystal lattice.

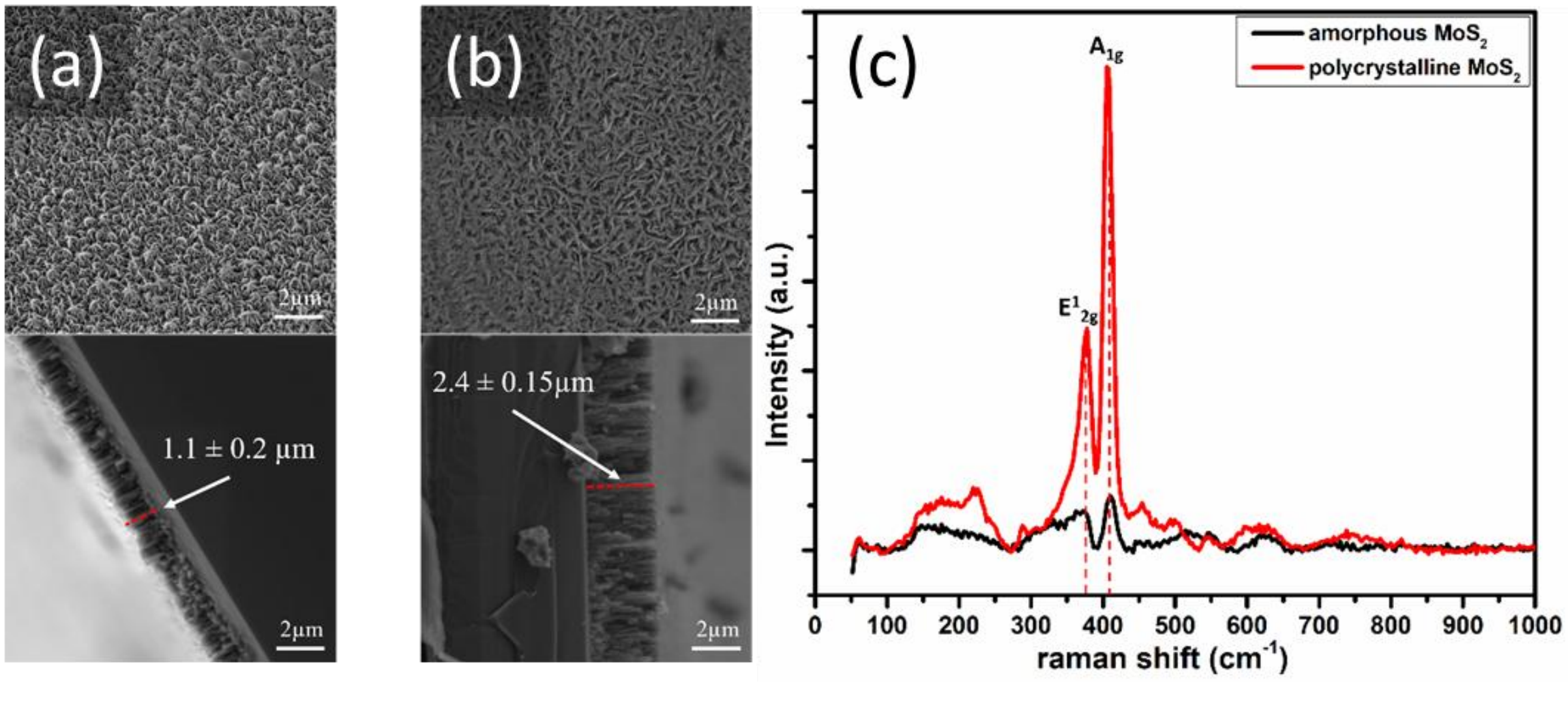

**Figure 1. (a)** Top view and cross section of the amorphous $Mo_2$ sample addressed in this study as imaged by SEM. **(b)** Same for the polycrystalline $Mo_2$. **(c)** Comparison of the Raman spectra measured on the two materials.

Figure 2 shows two AFM topographies of the amorphous and polycrystalline $MoS_2$ before scratching. The root mean square (RMS) surface roughness values for both materials are approximately150 nm and 90 nm, respectively. Both surfaces are isotropic with a correlation length of 0.28 µm and 0.15 µm respectively, as defined by the distance where the 2D autocorrelation of the surface topography decays to half of its peak value (not shown). While the first value corresponds to the characteristic size of the bulges on the top of the amorphous surface, the second value is consistent with the width of the randomly oriented needles forming the polycrystalline surface.

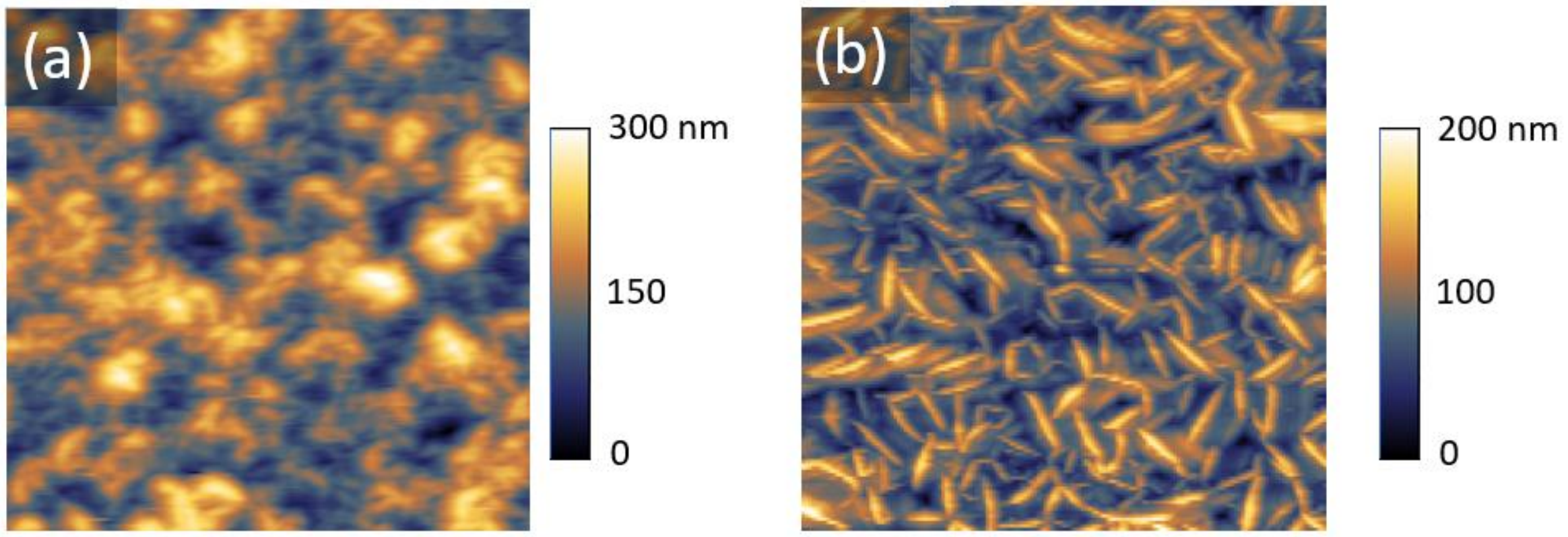

**Figure 2**. Representative AFM images of the pristine **(a)** amorphous and **(b)** polycrystalline $MoS_2$ surfaces. Frame sizes: 6×6 $\mu m^2$.

## 2.2. Nanoindentation characterization

Microscratching was performed by a G200 nanoindenter (KLA, Milpitas, CA, USA) with Nanosuite control software v. 6.30.457. The scratch tool was a three-sided diamond pyramid with an angle of 65.3° between the center line and the three faces. Such a so-called Berkovich tip was produced by Synton-MDP (Nidau, CH). Scratches were made along a distance of 60 µm in the direction of one of edges of the pyramid edges with four different loads, 0.1, 0.5, 1, and 2 mN, and four different speeds, 5, 50, 100, and 500 um/s.

## 2.3. Atomic force microscopy characterization

A commercial DriveAFM (Nanosurf, Liestal, Switzerland) was used for imaging the $MoS_2$ surfaces and also for scratching them with loading forces up to 15 µN. Both standard and off-resonance (ORT) tapping mode were used for imaging [10], with PPP-NCHR (Nano Sensors) and WM0.6 (Nanosurf) probes respectively. A diamond coated DT-NCHR probe (Nanosensors) with spring constant of 100 N/m was used for scratching. Wear tests and imaging were performed under ambient conditions (RH = 21%, T = 23 °C). Standard formulas of continuum mechanics were used to calibrate the forces acting on the rectangular [11] and V-shaped [12] probes. For AFM-based friction measurements, PPP-LFMR probes from Nanosensors, with a spring constant of 0.4 N/m were employed. The data were elaborated with Gwyiddion 2.63 [13].

## 3. Results and discussion

### 3.1. Microwear resistance of amorphous and polycrystalline $MoS_2$

The polycrystalline and amorphous $MoS_2$ films were first scratched with the Berkovich indenter along 16 straight lines divided into four groups according to the scheme in Fig. 3a. Within each group the scan speed $v$ was fixed at a value between 500 μm/s to 5 μm/s , and the normal force $F_N$ was varied between 100 μN and 2 mN. Figures 3b and 3b show representative wear tracks on each material, as imaged by AFM. Corresponding images for the remaining wear tracks are provided in the Supp. Mat.

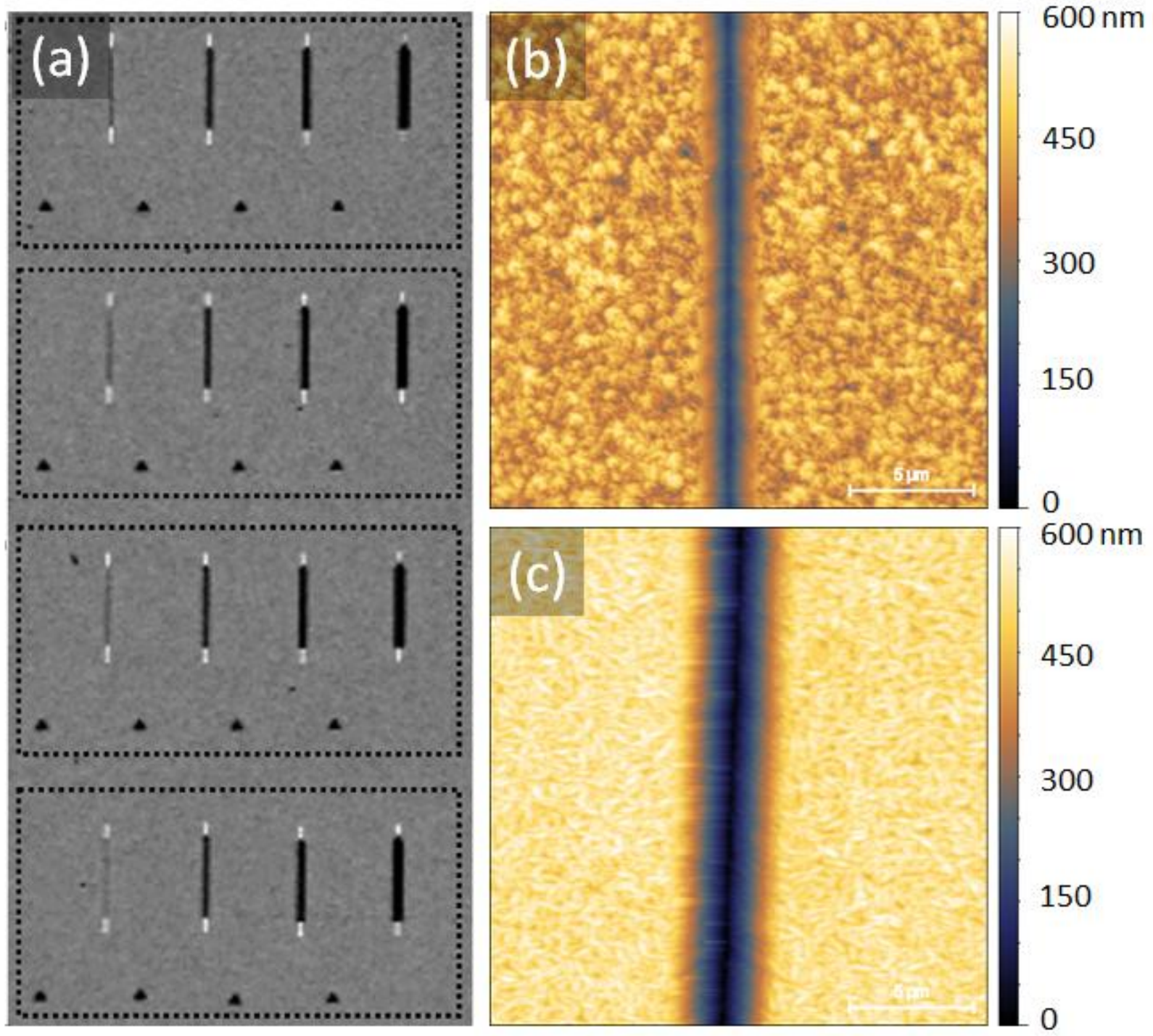

**Figure 3**. **(a)** Scratch pattern obtained with a Berkovich nanoindenter on amorphous $MoS_2$. The four panels correspond to a scan velocity $v$ = 500, 100, 50, and 5 μm/s (bottom to top). In each panel a loading force $F_N$ of 100, 500, 1000, and 2000 μN was applied (left to right). **(b, c)** Wear tracks on amorphous and polycrystalline $MoS_2$ for $v$ = 5 μm/s and $F_N$ = 500 nN as imaged by AFM. Frames sizes: 20×20 μm$^2$.

As shown in Fig. 4a and Fig. 4b the depth $h$ and the width $w$ of the wear tracks, as defined in Fig. 4c, were found to increase with the loading force $F_N$ on both polycrystalline (dashed lines) and amorphous $MoS_2$ (continuous lines), up to the maximum value of $F_N$ = 2 mN. In all cases the values of $h$ and $w$ were approximately double on polycrystalline $MoS_2$, suggesting that the wear rate on this material was about four times larger than that of amorphous $MoS_2$. Since the wear tracks reached a maximum depth of 2.4 μm and 1.1 μm respectively, i.e., about half of the film thickness in both cases, the substrate had no influence on the results. Moreover, we have also observed a slight increase of $h$ and $w$ with the scan speed $v$ on polycrystalline $MoS_2$, whereas no significant velocity dependence was observed on amorphous $MoS_2$ (with $v$ varying between 50 μm/s and 500 μm/s). All in all, these results suggest that polycrystalline $MoS_2$ is disrupted more easily than the amorphous form.

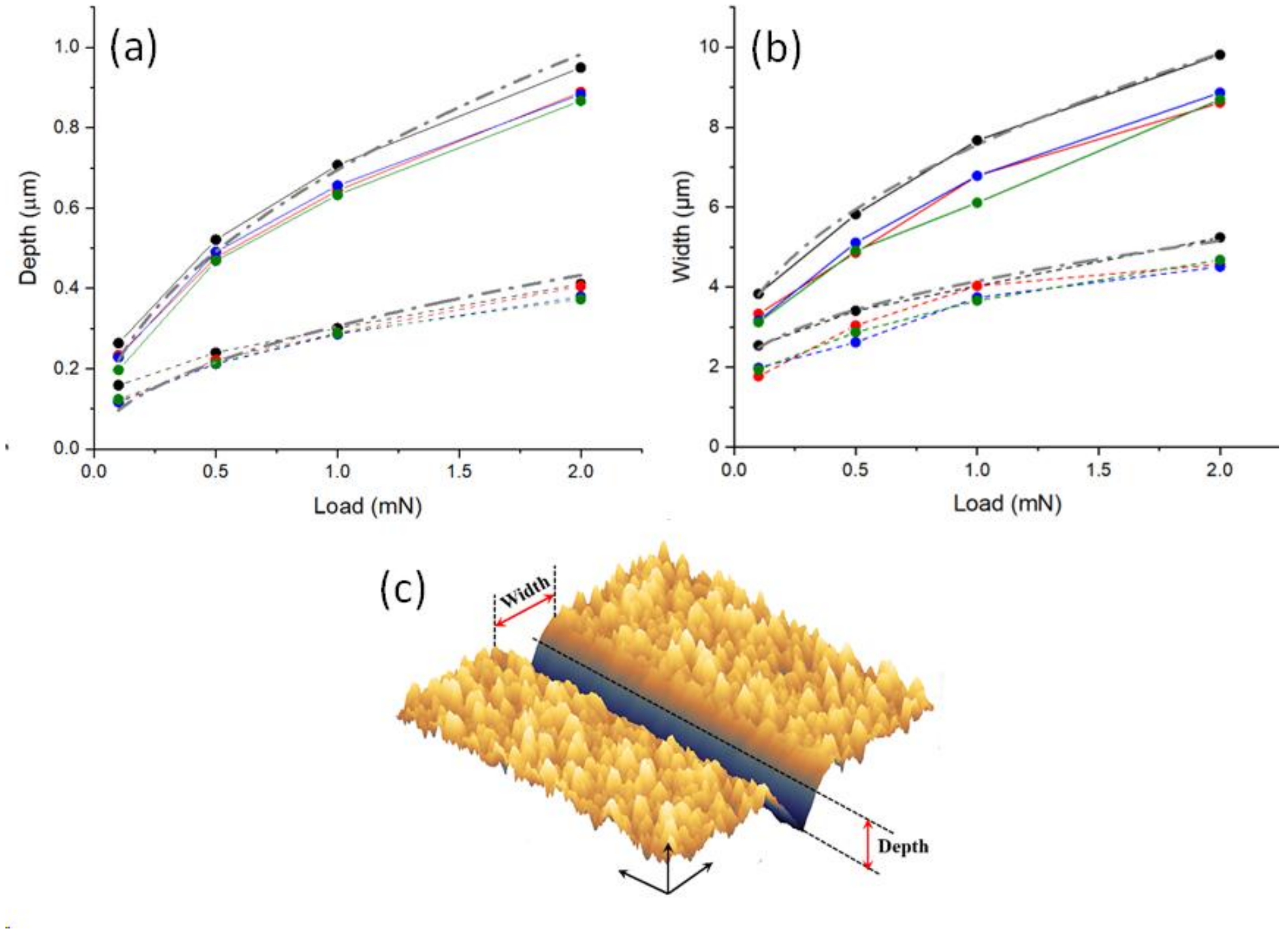

**Figure 4.** Load dependence at the mN level of **(a)** the scratch depth and **(b)** the scratch width on amorphous $MoS_2$ (continuous curves) and polycrystalline $MoS_2$ (dashed curves) with $v = 5$, 50, 100 and 500 µm/s. The dot-dashed curves correspond to the best fits obtained with the function $\sqrt{F_N}$ for the maximum scratch velocity. **(c)** 3D view of a representative scratch made by the Berkovich indenter.

To gain additional information on the hardness of the two films, we have fitted the experimental data with the well-known Archard wear model [14]. This model assumes that the area $A$ of an arbitrary cross-section of the wear groove is proportional to the normal force $F_N$ according to the relation $A = KF_N/H$, where $H$ is the hardness of the sample surface and $K$ is a dimensionless constant. Since the wear groove has a triangular section in our case: $A = wh/2 = h^2 \tan\alpha$, where $\alpha$ is half aperture angle of the groove, so that

$$h = C\sqrt{F_N}, \qquad \text{with} \quad C = \sqrt{\frac{K}{H \tan\alpha}} \quad . \tag{1}$$

The data in Fig. 4 are fully consistent with the trend given by eq. (1). From the fit of the points in Fig. 4(a) $C$ ranges between 280 and 306 µm/$\sqrt{\mathrm{N}}$ for amorphous $MoS_2$, and between 626 and 695 µm/$\sqrt{\mathrm{N}}$ for polycrystalline $MoS_2$. A similar trend, apart for an offset $w_0$, is obtained for the scratch width, meaning that the wear rate is approximately four times larger on polycrystalline $MoS_2$. This is not a trivial result, considering that the relation (1) was introduced as a phenomenological description for rough surfaces with multiasperity sliding contacts, and it has rarely been applied to the wear caused by singular tips [15]. The hardness has been estimated with the Oliver-Pharr method [16] applied to independent indentation measurements (see Supp. Mat.). It is $H = 0.483$

GPa for amorphous $MoS_2$ and $H = 0.221$ GPa for polycrystalline $MoS_2$. With the aforementioned values of $C$ and $\alpha \approx 5°$ and 8° for the two forms (as estimated from AFM topographies) we conclude that $K \approx 3.6$ and $K \approx 14$, respectively. Considering that $K$ is essentially a measure of the probability of debris detachment [14], the polycrystalline form appears much more friable than the amorphous one. A quantitative comparison is nevertheless made difficult by compaction effects affecting both materials during the scratching: please see the change in the RMS values of the surface roughness presented in Sec. 3.2.

The discussion would not be complete without mentioning a noticeable effect observed in the wear tracks. The shaded image in Fig. 5(a) shows the presence of transverse bands with the orientation of the Berkovich indenter used to scratch the polycrystalline $MoS_2$. These features have already been reported on silica glasses after scratching them with a similar tool [17]. The sharp edges of the bands observed in that case allowed us to recognize that the repetition distance between the bands increases linearly with the scan velocity, which was consistently attributed to the stick-slip motion of the nanoindenter. A similar conclusion is not straightforward in the present case since the transverse bands are barely recognizable. We can only state that, while driven at a constant speed, the indenter was possibly sped up and slowed down in an irregular fashion (due to local variations of the kinetic friction), resulting in a modulated shape of the wear track. The effect was also visible on the narrower tracks formed on the amorphous $MoS_2$ surface (Fig. 5b).

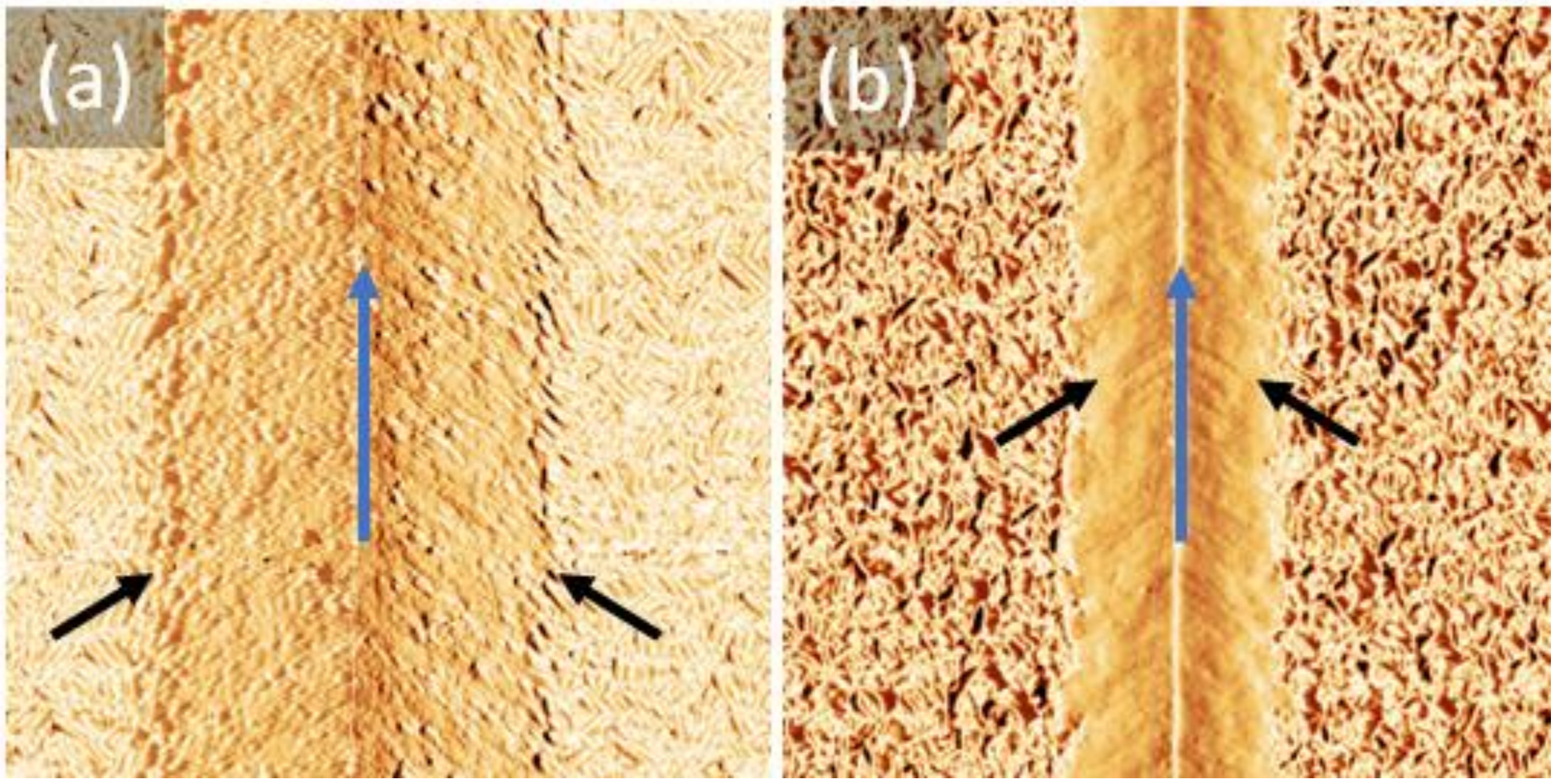

**Figure 5.** Shaded AFM topography corresponding to a scratch done **(a)** on the polycrystalline $MoS_2$ and **(b)** on the amorphous $MoS_2$ with a normal force $F_N = 2$ mN and a scratch velocity $v = 100$ nm. The blue arrows show the scratch direction. The black arrows highlight the orientation of the band structures observed in the wear track. Frame sizes: 16×16 μm$^2$.

### 3.2. Nanotribological response of amorphous and polycrystalline $MoS_2$

The estimation of the width and depth of the wear tracks turned out to be more difficult when the polycrystalline and amorphous $MoS_2$ surfaces were scratched by AFM (with a normal force $F_N$ up to 15 μN). With $F_N = 2$ μN (Fig. 6a) or below the width of the wear track on polycrystalline $MoS_2$ was indeed comparable to the width of the needles. On amorphous $MoS_2$ the wear track was barely visible when a force $F_N = 2$ μN was applied (Fig. 6b). Nevertheless, both width and depth of the wear track were found to increase with the applied load above this value on both materials (and even below it in the case of polycrystalline $MoS_2$), as seen in Fig. 6c and 6d. Also in this case the depth *w* of the wear track, when visible, was found to be about two times larger on the polycrystalline form. Since the width *w* was also larger, we have again a strong indication that the amorphous form is more resistant to ploughing wear. Due to the significant spread of the experimental data in Fig. 6a and 6b, an estimation of the relative hardness of the two materials based on the Archard model was not possible in this case.

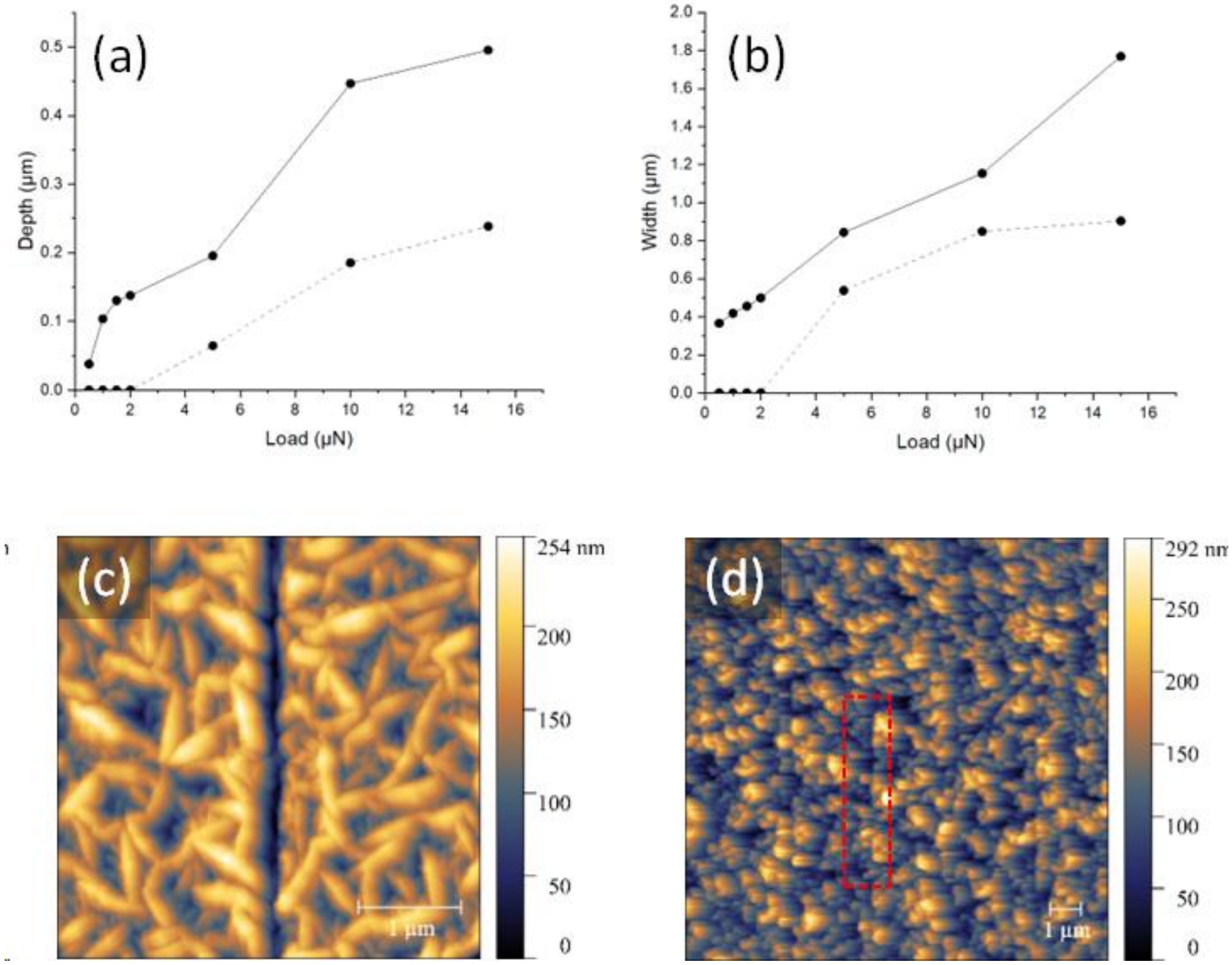

**Figure 6**. Load dependence of **(a)** the scratch depth and **(b)** the scratch width on amorphous $MoS_2$ (continuous curves) and polycrystalline $MoS_2$ (dashed curves) with $v$ = 5, 50, 100 and 500 µm/s as scraped by AFM. **(c)** AFM image a scratch made with 2 µN polycrystalline $MoS_2$. Frame size: 4×4 $\mu m^2$. **(d)** Same on amorphous $MoS_2$. Frame size: 12×12 $\mu m^2$. The red rectangle highlights the scratch barely visible in such case.

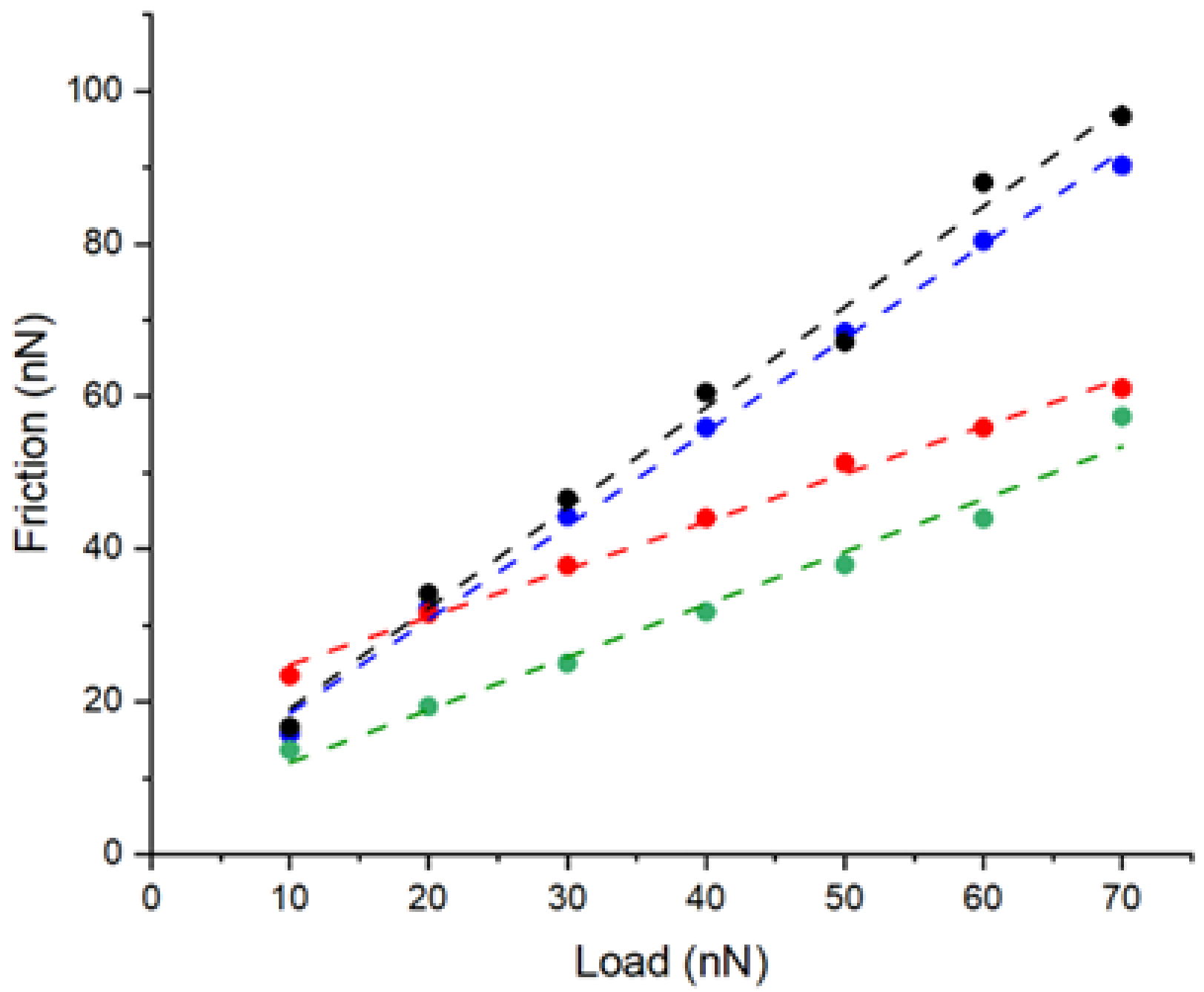

**Figure 7. (a)** Load dependence of the friction force on pristine and scratched regions of amorphous and polycrystalline $MoS_2$. The blue and black data points correspond to pristine polycrystalline and amorphous $MoS_2$. The green and red points correspond to the same materials after scratching. The dashed lines are linear fits.

As a final step, we have used AFM to characterize the lateral (friction) force $F_L$ acting on a sharp silicon tip sliding on amorphous and polycrystalline $MoS_2$ in their pristine conditions or after scratching them with the Berkovich indenter. The friction measurements were conducted with a scan speed of 5 μm/s and a normal force up to 70 nN. On the pristine surfaces, the average friction values were found to be similar on both materials (Fig. 7a). Friction increased linearly with $F_N$ in all cases with a friction coefficient of $\mu$ = 1.31 and 1.23 for the amorphous and polycrystalline form respectively (as estimated from the slopes of the fitting lines in the figure). On the scratched areas, the friction force was found to be higher on the polycrystalline material, whereas the friction coefficient remained comparable, and in both cases well below the values observed in the pristine areas ($\mu$ = 0.69 on amorphous vs. $\mu$ = 0.63 on polycrystalline $MoS_2$). This result is possibly due to surface smoothing caused by the pressure exerted by the tip while scratching, which can be visually appreciated in Fig. 5a-b. The high values of $\mu$ recorded on the pristine surfaces (both of them above 1) denote a strong influence of the surface morphology on the lateral force. In the wear track formed on the amorphous material the RMS roughness was indeed found to decrease by a factor 5 (from 54 nm to 11 nm). On the polycrystalline material it appears that the RMS value decreased by a factor 2 (from 24 nm to 12 nm), but, considering that the resolution of the porous sample is limited by the size of the AFM tip, this factor is possibly larger. It means that the interlocking between tip and surface is reduced by scratching, and sliding becomes smoother afterwards. Whether material properties were also modified during scratching and/or the concentration of environmental contaminants on the surfaces was changed cannot be assessed with the available set-ups. .

## 4. Conclusion

This research analyzed the surface characteristics and friction behavior of both polycrystalline and amorphous $MoS_2$ during the process of scratching. The study involved the use of an indenter and an atomic force microscope to scratch a 2.4 μm thick polycrystalline film and a 1.1 μm thick amorphous film. After making a scratch, depth, width, and friction force were examined.

For indenter-based scratching, a range of forces (0.1 to 2 mN) and speeds (5 to 500 μm/s) were used. The comparison of the scratch depth and width revealed a significant difference in the deformation between the polycrystalline and amorphous materials. Under all examined conditions, the wear grooves on polycrystalline $MoS_2$ were approximately two times deeper and wider than on amorphous $MoS_2$. The results also showed that, in both cases, the wear volume increases with the normal load consistently with the Archard law. On the other hand, the scratch speed was found to have only a weak influence on polycrystalline $MoS_2$, and to be almost irrelevant on the amorphous form. Regarding friction, the comparison between pristine and scratched regions on the two materials revealed a linear dependence on the normal load in all cases, with a significant reduction of the COF in the regions made smoother by scratching. A comparison of independent nanoindentation tests also allows us to conclude that the polycrystalline form is more friable than the amorphous one, and, therefore, less suited as a solid coating in ambient conditions.

## Acknowledgments

The support of the Polish National Science Center (NCN) via the Opus Grant No. UMO/2021/43/B/ST5/00705 is gratefully acknowledged by all authors. HK and EG are also thankful to the Strategic Program Excellence Initiative at the Jagiellonian University SciMat (Grant No. U1U/P05/NO/01.05). N.D. acknowledges support by the European Union under the project Robotics and advanced industrial production (reg. no. CZ.02.01.01/00/22_008/0004590). T.P. acknowledges the support of the Czech Science Foundation (project No. 23-07785S).

# Supplementary material

Fig. S1 and Fig. S2 provide a detailed overview of 16 scratches made by the indenter on the amorphous and polycrystalline materials, respectively. The presentation is structured in a 4x4 matrix format, highlighting the force parameters and scanning speed. The force parameter varies in increments of 0.1, 0.5, 1, and 2 mN across the columns, whereas the scanning speed progresses within each row, ranging from 5 to 50, 100, and 500 µm/s . The imaging process was carried out using an atomic force microscope in tapping mode. We used a PPP-NCHR probe to observe the surface features with highest possible resolution.

Fig. S3 shows a collection of images captured by the atomic force microscope, which displays scratches on two different materials. The first row shows scratches on amorphous material, while the second row shows scratches on polycrystalline material. Within each row, the force used to create the scratches was increased from 2 to 15 µN . The DT-NCHR cantilever was used to create the scratches and capture the images, allowing for a detailed examination of the surface features and material properties.

Fig. S4 demonstrates how to create a scratch using the indenter. The process begins with the Berkovich tip making initial contact with the material's surface, as shown by the triangle in the figure. The scratch operation then begins from a short distance away from the initial contact point. The figure provides a visual representation of the key stages in the scratching process, including a detailed explanation of how the scratch is initiated.

Fig. S5 shows two representative load vs. displacement curves (out of 11 curves) as obtained with the nanoindenter. The data were further analyzed using the Oliver-Pharr method and the hardness resulting on the two samples was averaged to provide the values reported in the text.

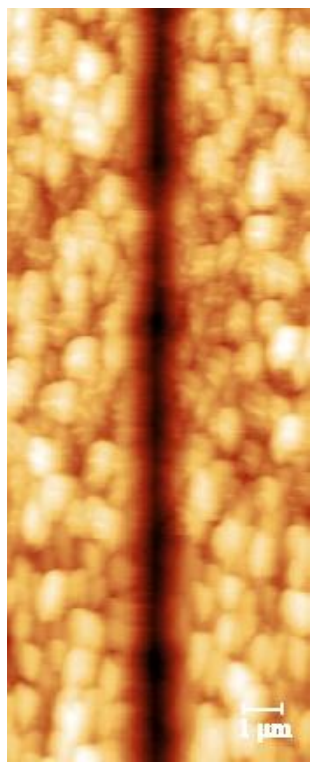

V= 5 µm/s, F= 0.1 mN

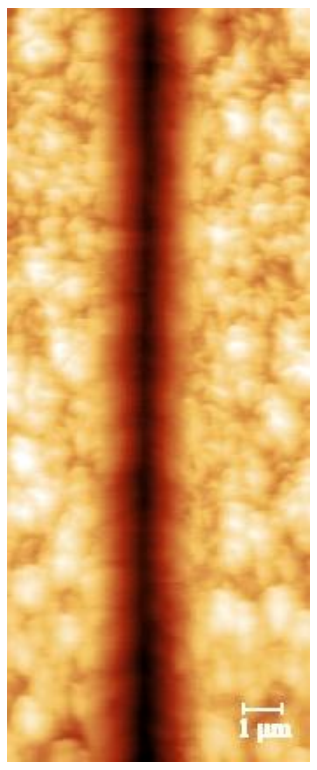

V= 5 µm/s, F= 0.5 mN

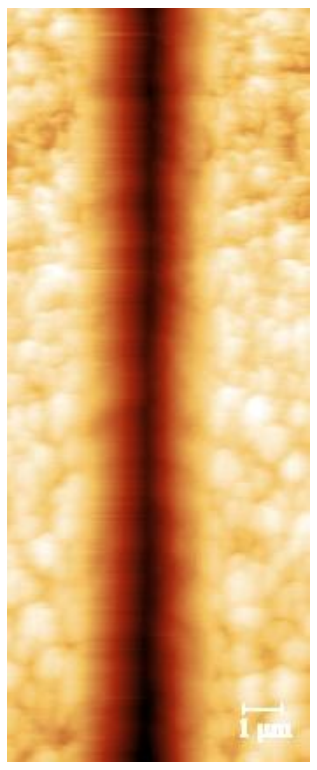

V= 5 µm/s, F= 1 mN

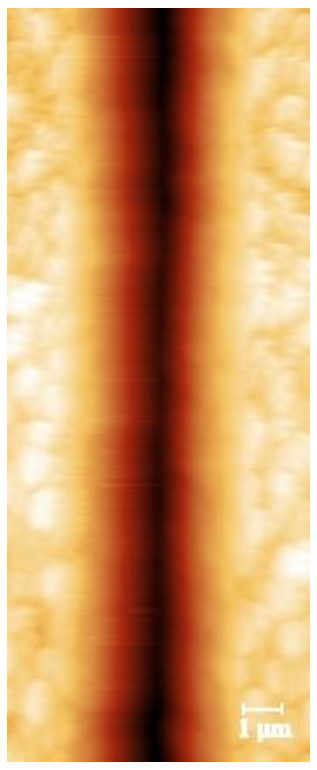

V= 5 µm/s, F= 2 mN

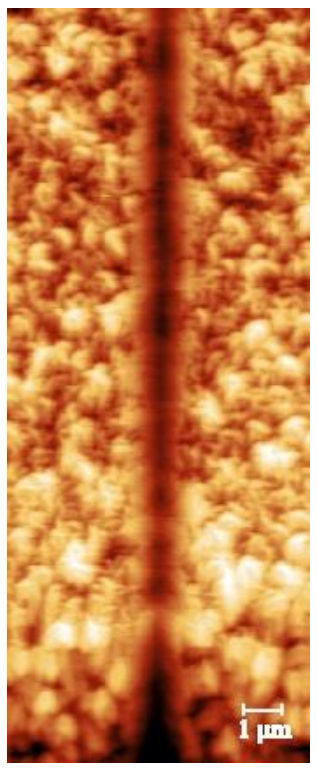
V= 50 µm/s, F= 0.1 mN

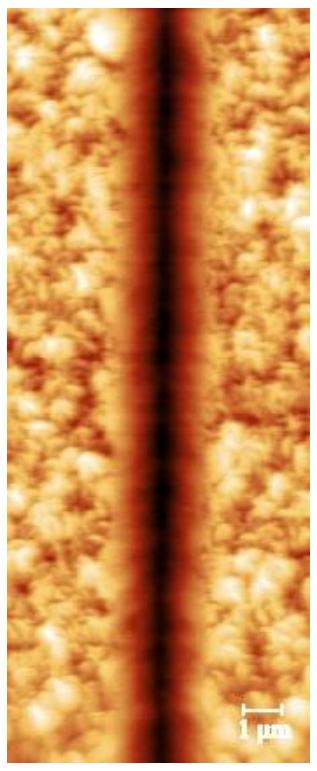
V= 50 µm/s, F= 0.5 mN

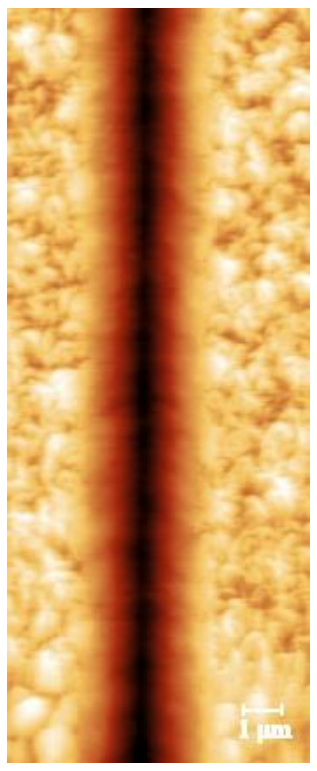
V= 50 µm/s, F= 1 mN

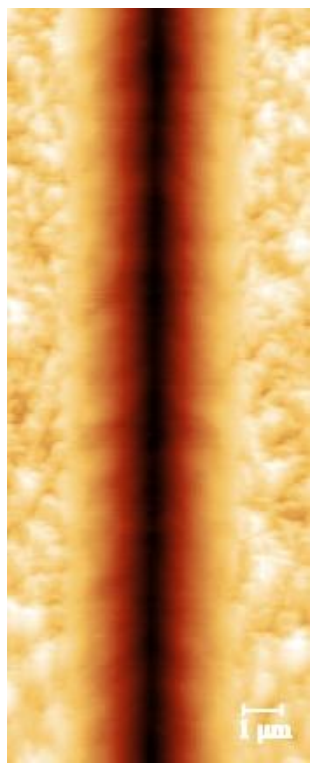
V= 50 µm/s, F= 2 mN

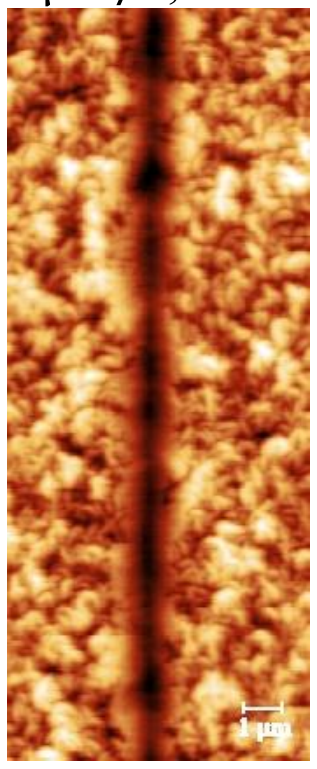
V= 100 µm/s, F= 0.1 mN

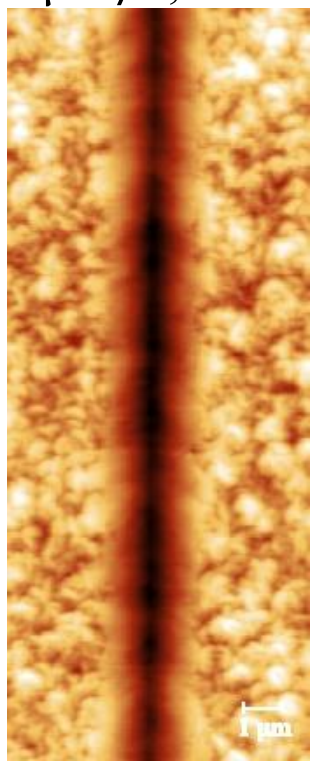
V= 100 µm/s, F= 0.5 mN

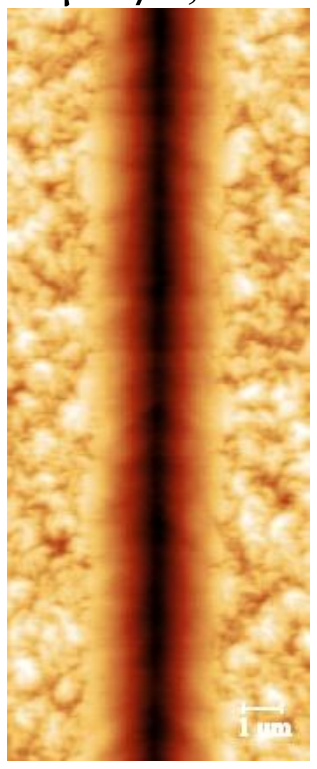
V= 100 µm/s, F= 1 mN

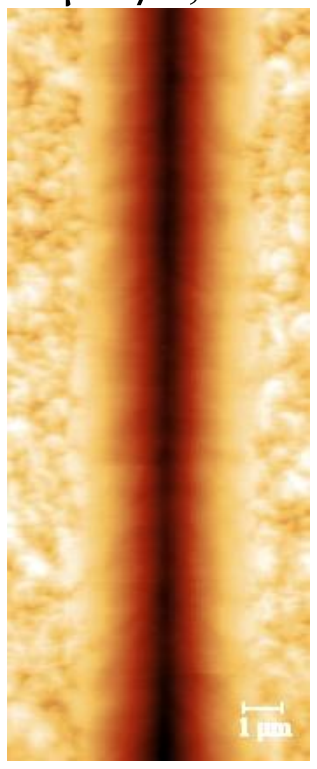
V= 100 µm/s, F= 2 mN

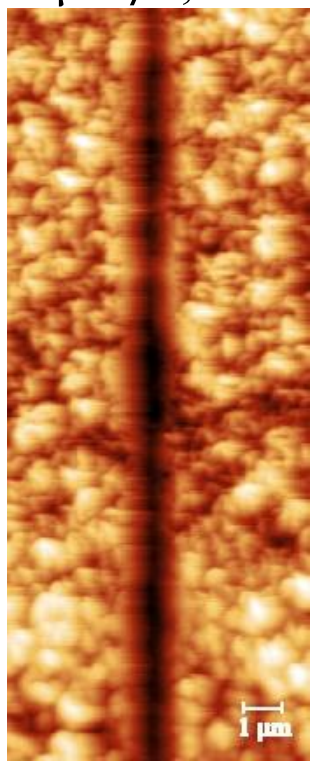
V= 500 µm/s, F= 0.1 mN

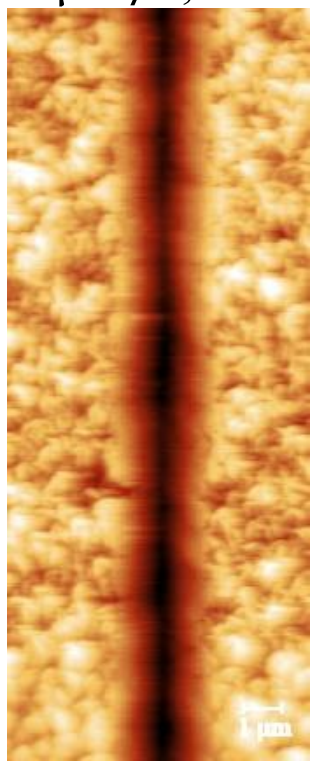
V= 500 µm/s, F= 0.5 mN

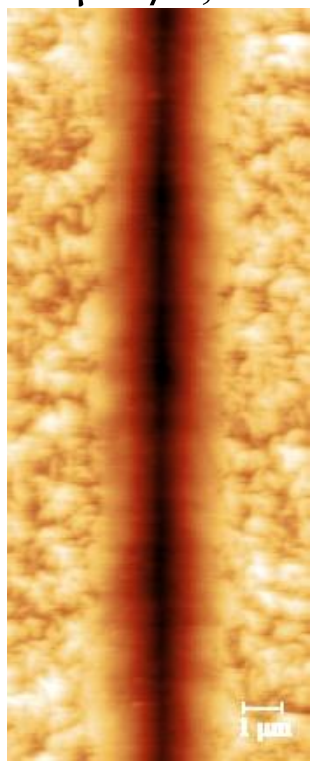
V= 500 µm/s, F= 1 mN

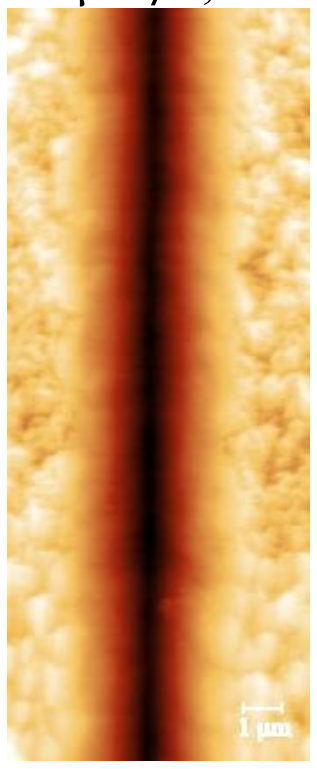
V= 500 µm/s, F= 2 mN

**Figure S1**. AFM image of the scratched area with an indentor, amorphous (Tapping mode image)

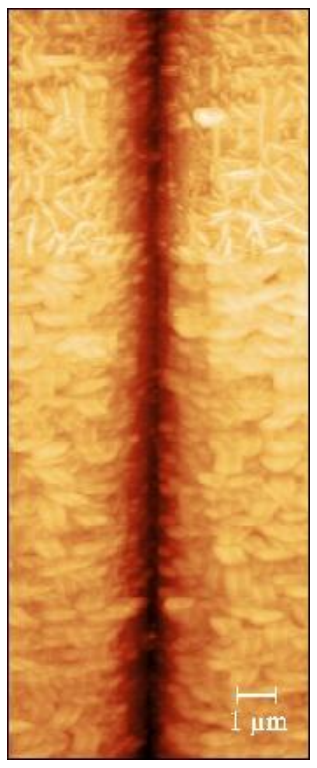
V= 5 µm/s, F= 0.1 mN

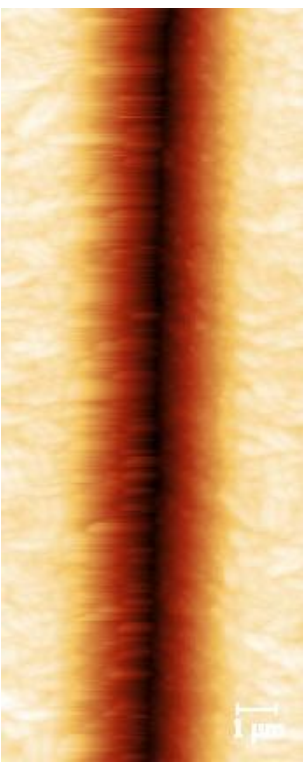
V= 5 µm/s, F= 0.5 mN

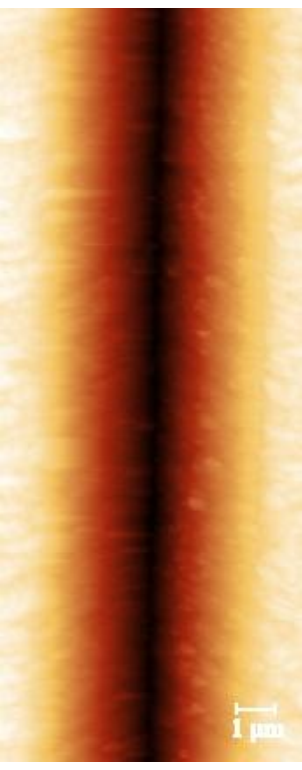
V= 5 µm/s, F= 1 mN

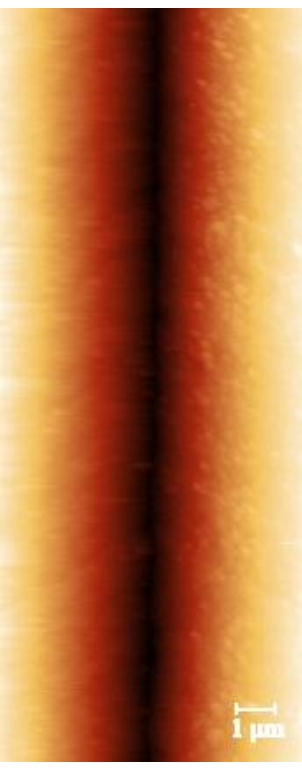
V= 5 µm/s, F= 2 mN

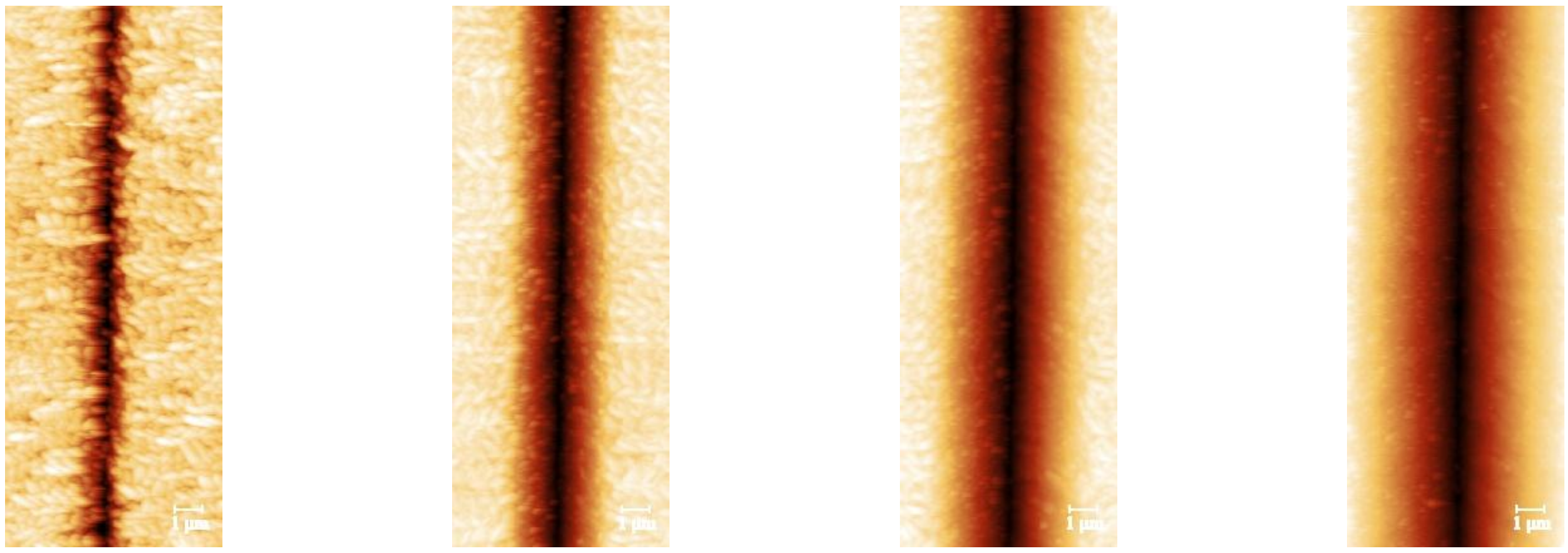

V= 50 µm/s, F= 0.1 mN   V= 50 µm/s, F= 0.5 mN   V= 50 µm/s, F= 1 mN   V= 50 µm/s, F= 2 mN

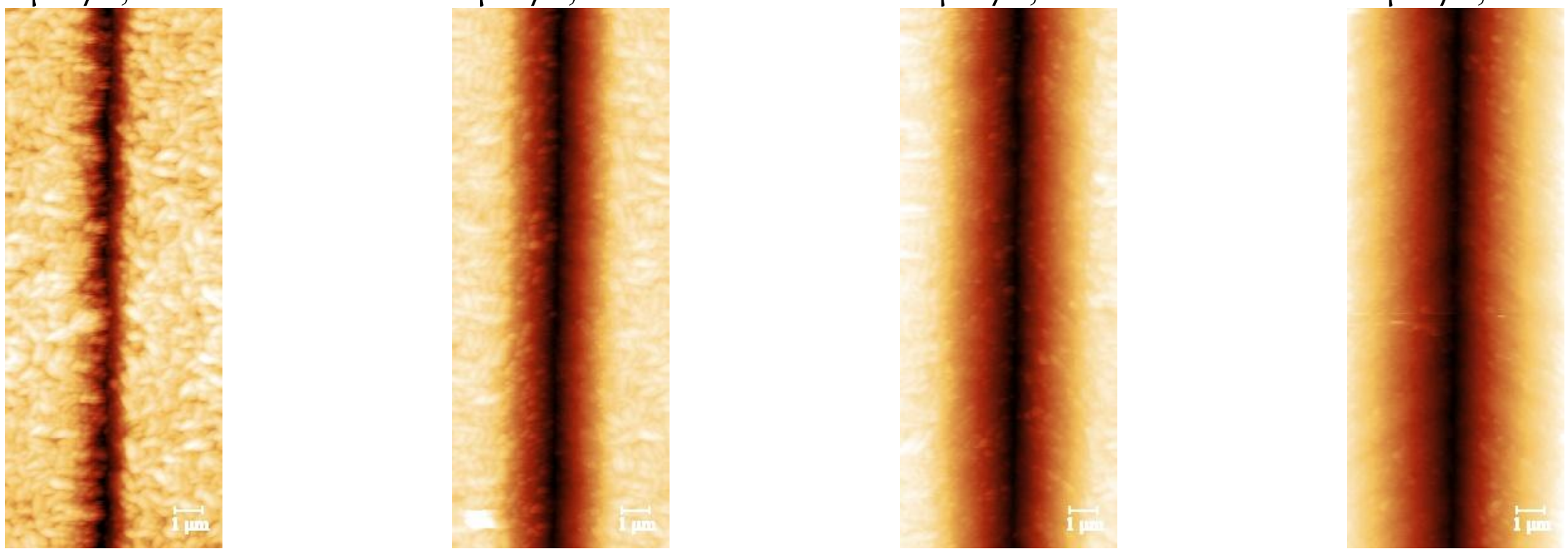

V= 100 µm/s, F= 0.1 mN   V= 100 µm/s, F= 0.5 mN   V= 100 µm/s, F= 1 mN   V= 100 µm/s, F= 2 mN

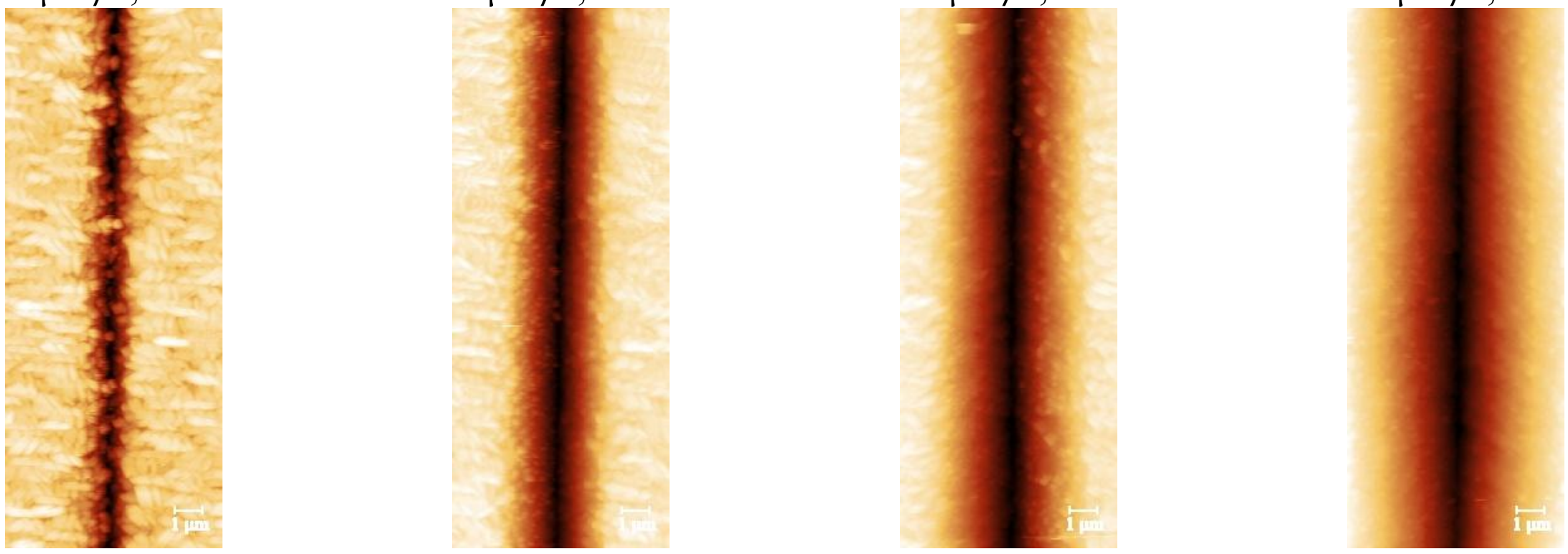

V= 500 µm/s, F= 0.1 mN   V= 500 µm/s, F= 0.5 mN   V= 500 µm/s, F= 1 mN   V= 500 µm/s, F= 2 mN

**Figure S2**. AFM image of the scratched area with an indentor, polycrystalline (Tapping mode image)

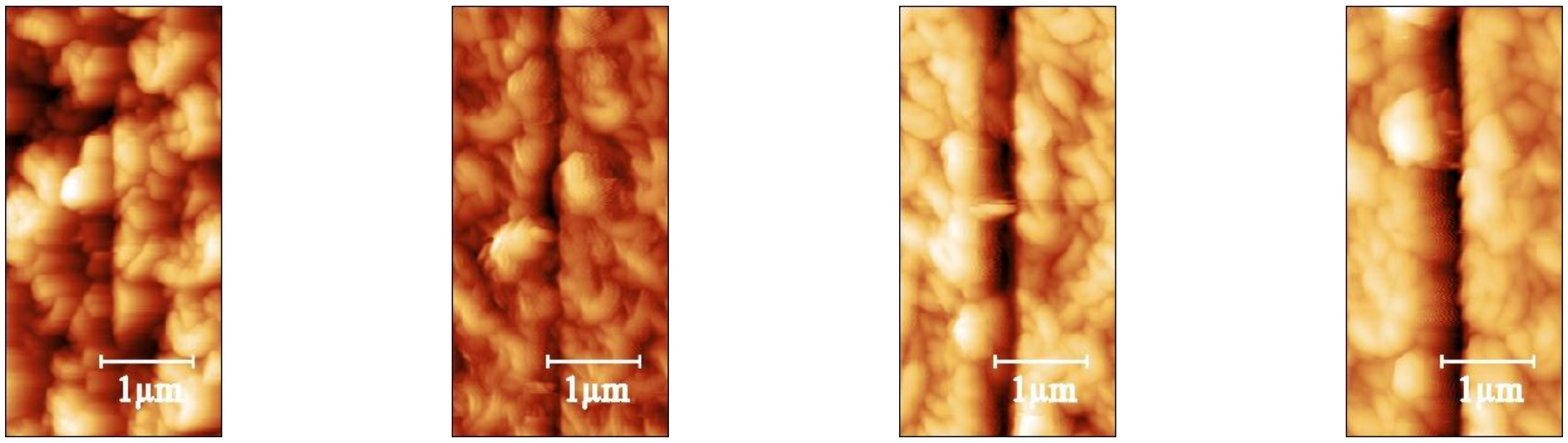

F= 2 µN/s   F= 5 µN/s   F= 10 µN/s   F= 15 µN/s

Amorphous

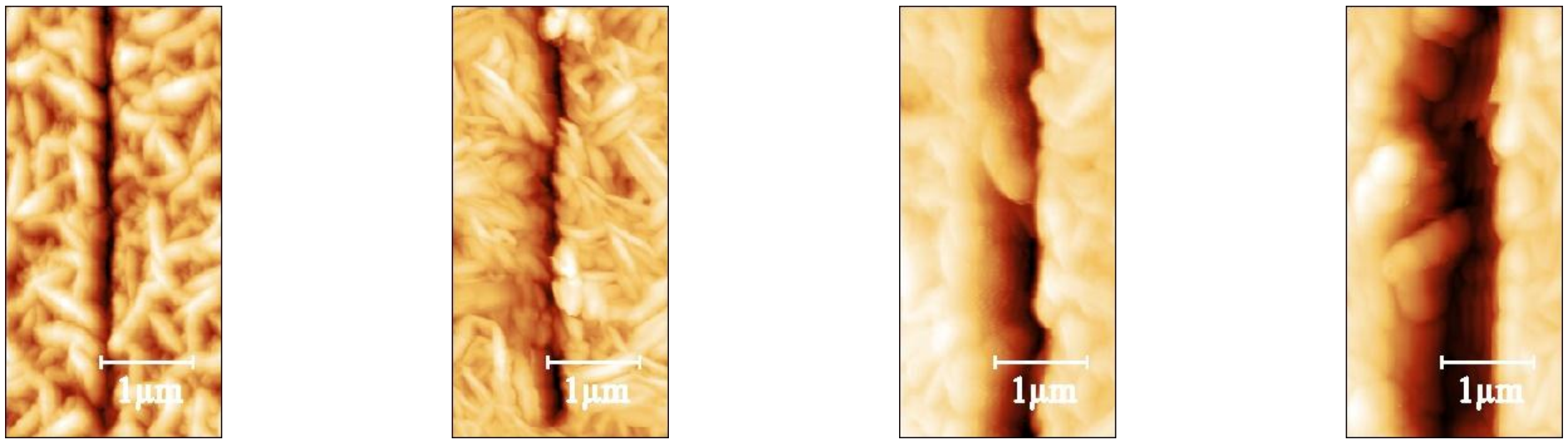

F= 2 µN/s F= 5 µN/s F= 10 µN/s F= 15 µN/s

Polycrystalline

**Figure S3**. AFM image on the scratched area with the AFM (contact mode image)

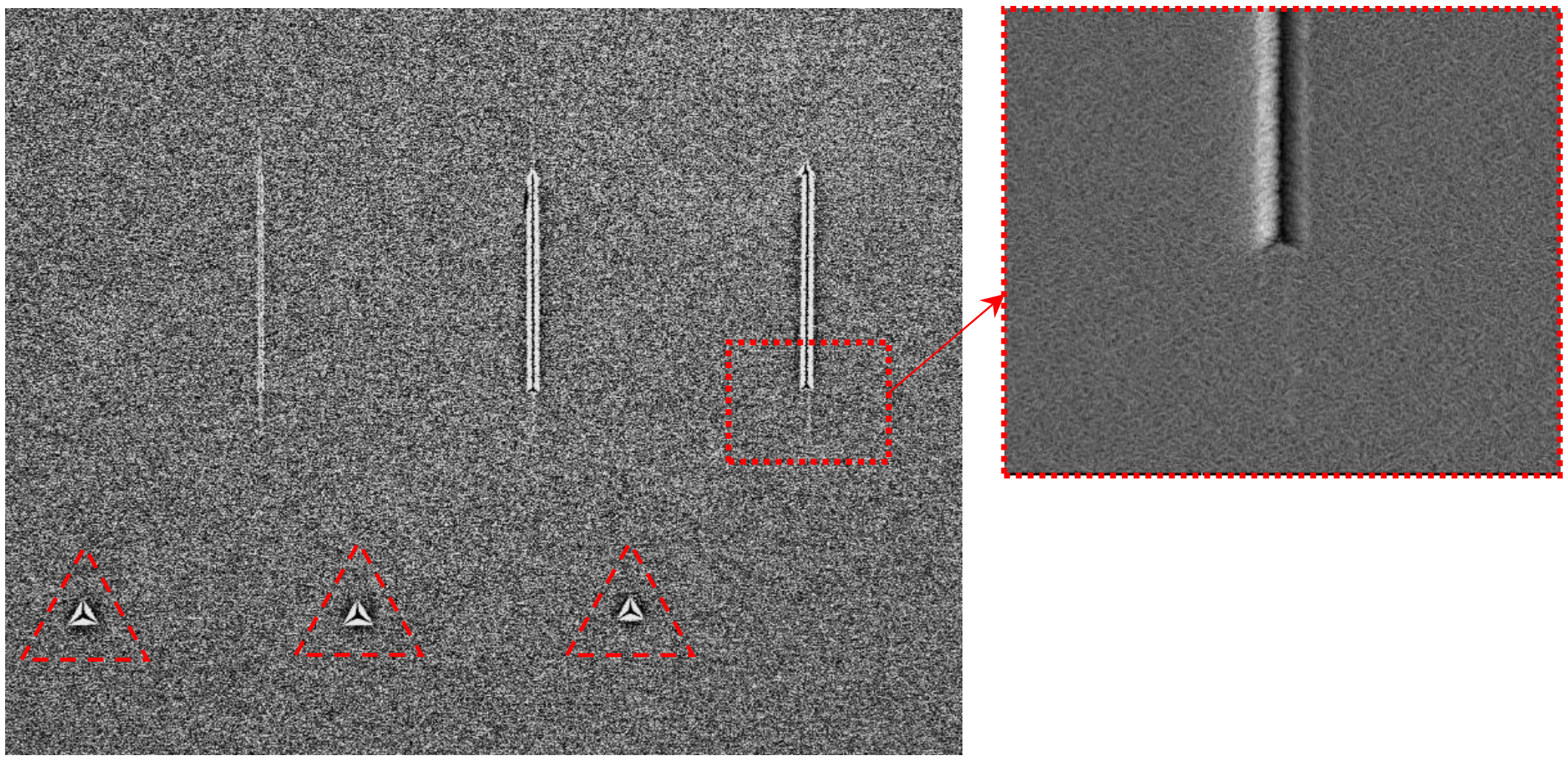

**Figure S4**. Scratching with indenter

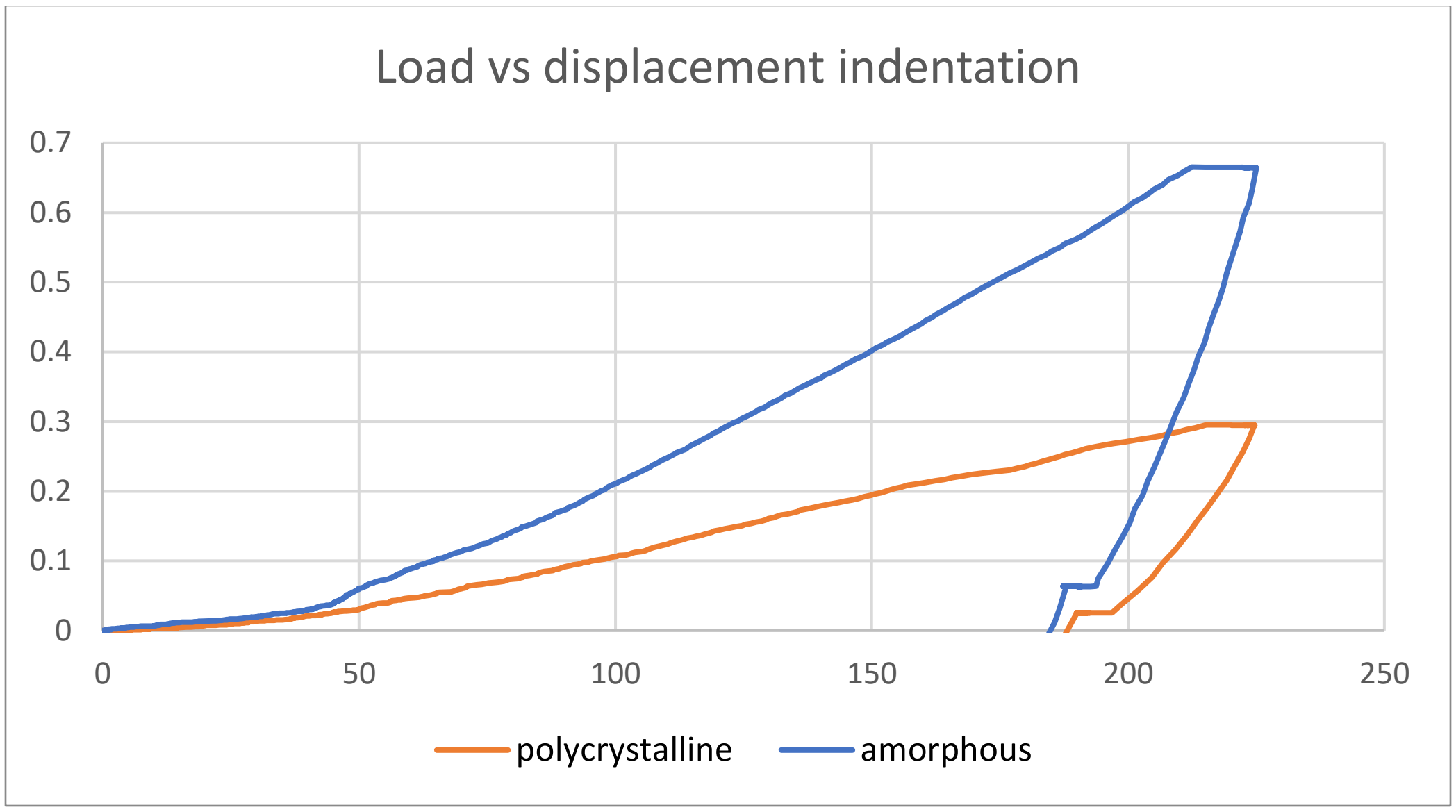

**Figure S5**. Representative load-displacement curves recorded on polycrystalline and amorphous $MoS_2$ using the nanoindenter. The units for the vertical and horizontal axes are mN and nm respectively.